\documentclass[nofootinbib,prd,twocolumn,showpacs,showkeys,preprintnumbers]{revtex4-1}
\usepackage{hyperref,amssymb,amsmath,mathrsfs,bm,graphicx}
\begin{document}
\title {Shearing and geodesic axially symmetric perfect  fluids that do not produce  gravitational radiation}
\author{L. Herrera}
\email{lherrera@usal.es}
\affiliation{Escuela de F\'\i sica, Facultad de Ciencias, Universidad Central de Venezuela, Caracas 1050, Venezuela and Instituto Universitario de F\'isica
Fundamental y Matem\'aticas, Universidad de Salamanca, Salamanca 37007, Spain}
\author{A. Di Prisco}
\email{alicia.diprisco@ciens.ucv.ve}
\affiliation{Escuela de F\'\i sica, Facultad de Ciencias, Universidad Central de Venezuela, Caracas 1050, Venezuela}
\author{J. Ospino}
\email{j.ospino@usal.es}
\affiliation{Departamento de Matem\'atica Aplicada and Instituto Universitario de F\'isica
Fundamental y Matem\'aticas, Universidad de Salamanca, Salamanca 37007, Spain}
\author{J.Carot }
\email{jcarot@uib.cat}
\affiliation{Departament de  F\'{\i}sica, Universitat Illes Balears, E-07122 Palma de Mallorca, Spain}
\date{\today}
\begin{abstract}
Using a framework based on the $1+3$ formalism we carry out a study on  axially and reflection symmetric perfect  and geodesic fluids, looking for possible models of sources radiating gravitational waves. Therefore,  the fluid should be necessarily shearing, for otherwise the magnetic part of the Weyl tensor vanishes, leading to a vanishing of the super--Poynting vector. However, for the family of perfect, geodesic fluids considered here, it appears that all possible cases reduce to conformally flat, shear--free, vorticity--free, fluids,  i.e Friedmann-Roberston-Walker.  The super-Poynting vector vanishes and therefore  no gravitational radiation is expected to be produced.  The physical meaning of the obtained result is discussed.
\end{abstract}
\maketitle

\section{Introduction}
In a recent paper \cite{1} the $1+3$ formalism \cite{21cil, n1, 22cil, nin} has been used to develop a general framework for studying axially symmetric dissipative fluids. Besides some results exhibited in \cite{1}, the above mentioned framework has been  applied to the shear-free case \cite{shearfree}. As the result of such a study,  it follows that all geodesic and shear-free fluids, are irrotational, and as consequence of this, also purely electric. Such a result holds for a general fluid (not necessarily perfect). 

Now, if we define a state of intrinsic
gravitational radiation (at any given point), to be one in
which  the super-Poynting vector does not vanish for any  unit timelike vector   \cite{bel, agp, Mac}, then since the vanishing of the magnetic part of the Weyl tensor implies the vanishing of the super-Poynting vector, it is clear that  when looking for gravitationally radiating sources (at least under the geodesic condition) we should consider shearing fluids. It is worth recalling that the tight link between the super-Poynting vector and the existence of a state of radiation is firmly supported by the relationship between the former and the Bondi news function \cite{7} (see \cite{why} for a discussion on this point).

From the comments above, and as a further step towards the understanding of  gravitationally radiating sources, we shall consider in this work the simplest fluid distribution which we might believe to be  compatible with a non-vanishing super-Poynting vector, namely: perfect fluid under the geodesic condition.

However, our investigation shows that  all possible models, sourced by a perfect fluid (which of course includes the  pure dust configuration as a particular subcase), belonging to the family of the line element considered here, do not radiate gravitational waves during their evolution.

We shall discuss about this result and its relationship with the fact that the process of radiation (including absorption and/or Sommerfeld type conditions) is not a reversible one.

We shall heavily rely on the general framework developed in \cite{1}, keeping the same notation, and just reducing the general equations to the particular case considered here.  These will be presented in an Appendix.

\section{The perfect, geodesic, fluid}
We shall consider  axially and reflection symmetric perfect fluid distributions (not necessarily bounded). For such a system, we assume that  the  line element may be written in ``Weyl spherical coordinates'',  as:

\begin{equation}
ds^2=-A^2 dt^2 + B^2 \left(dr^2
+r^2d\theta^2\right)+C^2d\phi^2+2Gd\theta dt, \label{1b}
\end{equation}
where $A, B, C, G$ are positive functions of $t$, $r$ and $\theta$. We number the coordinates $x^0=t, x^1=r, x^2= \theta, x^3=\phi$.

The specific form of the line (\ref{1b}) deserves some comments. Our original goal, here as well as in \cite{1} and \cite{shearfree}, has been to describe the gravitational radiation process, through the physical properties of its source. Such an endeavour, should, eventually, lead to the obtention of a specific source of gravitational radiation.  With this purpose in mind, and in order to render the problem under consideration, analytically handable, we have imposed the highest degree of symmetry compatible with the existence of gravitational radiation. For vacuum, it is represented by the Bondi metric \cite{7}. Here, following the framework developed in \cite{1}, we have restricted the line element as much as possible, always allowing for the existence of gravitational radiation (at least for the most general matter distribution). As the result of these restrictions, we are able to manipulate the resulting expressions, by purely analytical procedures. However, this is obtained at the price of dealing with a space--time, that is not the more general one, compatible with axial symmetry (see the discussion below (\ref{nu1})).

The energy momentum tensor in the ``canonical'' form reads:
\begin{eqnarray}
{T}_{\alpha\beta}&=& (\mu+P) V_\alpha V_\beta+P g _{\alpha \beta} ,
\label{6bis}
\end{eqnarray}
where as usual, $\mu, P$ and  $V_\beta$ denote the energy density, the isotropic pressure, and  the four velocity, respectively. As in \cite{1} we are assuming the fluid to be comoving in the coordinates of (\ref{1b}).

The shear tensor is defined by two scalar functions $\sigma_I, \sigma_{II}$, which in terms of the metric functions read (see eqs.(20-25) in \cite{1}):

\begin{eqnarray}
2\sigma_I+\sigma_{II}&=&\frac{3}{A}\left(\frac{\dot B}{B}-\frac{\dot C}{C}\right),\label{primers}\\
2\sigma_{II}+\sigma_I&=&\frac{3}{A^2B^2r^2+G^2}\,\left[AB^2r^2\left(\frac{\dot B}{B}-\frac{\dot C}{C}\right)\right.\nonumber\\
&
&\left.+\frac{G^2}{A}\left(-\frac{\dot A}{A}+\frac{\dot G}{G}-\frac{\dot C}{C}\right)\right],\label{sigmas}
\end{eqnarray}
where dots denote derivatives with respect to $t$.

For the other kinematical variables (the expansion, the four acceleration and the vorticity) we have:

the expansion

\begin{eqnarray}
\Theta&=&V^\alpha_{;\alpha}\nonumber\\
&=&\frac{AB^2}{r^2A^2B^2+G^2}\,\left[r^2\left(2\frac{\dot B}{B}+\frac{\dot C}{C}\right)\right.\nonumber\\
&&+\left.\frac{G^2}{A^2B^2}\left(\frac{\dot B}{B}-\frac{\dot A}{A}+\frac{\dot G}{G}+\frac{\dot C}{C}\right)\right].
\label{theta}
\end{eqnarray}

The four acceleration
\begin{equation}
a_\alpha=V^\beta V_{\alpha;\beta}=a_I K_\alpha+a_{II}L_\alpha,\label{acc}
\end{equation}
\noindent with vectors $\bold K$ and $\bold L$  having components:
\begin{equation}
K_\alpha=(0, B, 0, 0);\qquad L_\alpha=(0, 0, \frac{\sqrt{A^2B^2r^2+G^2}}{A}, 0),\label{vec}
\end{equation}

\noindent and where the  two scalar functions ($a_I, a_{II}$) are defined by (see eq.(17) in \cite{1})
 \begin{eqnarray}
 a_I&=& \frac {A^{\prime} }{AB},
 \\
 a_{II}&=&\frac{A}{\sqrt{A^2B^2r^2+G^2}}\left[\frac{G}{A^2}\left(-\frac {\dot A}{A}+\frac {\dot G}{G}\right)+\frac {A_{,\theta}} {A}\right],
\label{acc'}
\end{eqnarray}
whereas the vorticity vector is defined through a single scalar $\Omega$, given by (see eq.(29) in \cite{1})

\begin{equation}
\Omega =\frac{(AG^\prime-2GA^\prime)}{2AB\sqrt{A^2B^2r^2+G^2}}\label{omega},
\end{equation}
where  primes  denote derivatives with respect to $r$.

\noindent We shall  restrict our system to the case of vanishing four--acceleration $a_\alpha=0$.
\noindent This  condition  implies that

\begin{equation}
a_I=\frac{A^\prime}{AB}=0\,\,\,\Rightarrow \,\,\, A=\tilde A(t,\theta),\label{a1}
\end{equation}
\noindent and
\begin{equation}
a_{II}=0\Rightarrow\frac{G}{A^2}(-\frac{\dot A}{A}+\frac{\dot G}{G})+\frac{A_{,\theta}}{A}=0,\label{a2}
\end{equation}
\noindent or

\begin{equation}
\left (\frac{G}{A}\right  )^{\bf{.}}=-A_{,\theta}\,\,\,\Rightarrow \,\, \frac{G}{A}=-\int A_{,\theta}dt+\tilde G(r,\theta).\label{a2'}
\end{equation}
\noindent Given that $ G(t,0,\theta)=0$, from (\ref{a2'}) we find that
\begin{equation}
A_{,\theta}=0 \,\,\,\Rightarrow\,\,\, A=\tilde A(t)\,\,\, {\rm and} \,\,\, G=\tilde G \tilde A.
\end{equation}

\noindent In this case, reparametrizing the time coordinate, the line element takes the form
\begin{equation}
ds^2=-dt^2+B^2\left  (dr^2+r^2d\theta ^2\right)+2\tilde G(r,\theta)dt d\theta+C^2d \phi ^2,\label{metric2}
\end{equation}

\noindent and the kinematical quantities become
\begin{equation}
\Theta=\frac{2B^2r^2+\tilde G^2}{B^2r^2+\tilde G^2}\frac{\dot B}{B}+\frac{\dot C}{C},\label{Thetaa}
\end{equation}

\begin{equation}
\Omega=\frac{\tilde G^\prime}{2 B\sqrt{B^2r^2+\tilde G^2}},\label{omegaa}
\end{equation}
\begin{equation}
\sigma _I=\frac{B^2r^2+2\tilde G^2}{B^2r^2+\tilde G^2}\frac{\dot B}{B}-\frac{\dot C}{C},\label{sigmaIa}
\end{equation}
\begin{equation}
\sigma_I-\sigma_{II}=\frac{3\tilde G^2}{B^2r^2+\tilde G^2}\frac{\dot B}{B}.\label{sigmasa}
\end{equation}
Now, from the regularity conditions, necessary to ensure elementary flatness in the vicinity of  the axis of symmetry, and in particular at the center (see \cite{1n}, \cite{2n}, \cite{3n}), we should require  that as $r\approx 0$
\begin{equation}
\Omega=\sum_{n \geq1}\Omega^{(n)}(t,\theta) r^{n},
\label{sum1}
\end{equation}
implying, because of (\ref{omegaa}) that in the neighborhood of the center
\begin{equation}
\tilde G=\sum_{n\geq 3}\tilde G^{(n)}(\theta) r^{n}.
\label{sum1}
\end{equation}
This last result in turn implies that as $r$ approaches $0$,
\begin{equation}
 \sigma_I- \sigma_{II}=\sum_{n\geq 4}\left[\sigma_I^{(n)}(t,\theta)-\sigma_{II}^{(n)}(t,\theta) \right]r^{n}.
\label{ref1}
\end{equation}

Now, for the length of an orbit at $t, \theta$ constant, to be $2\pi r$, close to the origin (elementary flatness), we may write, as $r\rightarrow 0$, 
\begin{equation}
C\approx r\gamma(t,\theta), 
\label{nuev1}
\end{equation}
implying
\begin{equation}
C^\prime \approx \gamma(t,\theta), \;\; C_{,\theta}\approx r\gamma_{,\theta},
\label{nuev3}
\end{equation}
where $\gamma(t,\theta)$ is an arbitrary function of its arguments, which as appears evident from the elementary flatness condition, cannot vanish anywhere within the fluid distribution.

Finally,  observe that a combination of (\ref{Thetaa})--(\ref{sigmasa}) produces 
\begin{equation}
(\sigma_I-\sigma_{II})(Br^2+\tilde G^2)=\tilde G^2(\Theta+\sigma_I).
\label{une1}
\end{equation}

We shall next make use of the full set of equations  deployed in \cite{1}, written for the specific case considered here, they are given in the Appendix A.

Thus, taking the time derivative of the above equation and combining  with (\ref{ecc1Y}), (\ref{ecc2Y}) and (\ref{ecc3Y}), we obtain
\begin{equation}
{\cal E}_{II}-{\cal E}_{I}=\frac{\sigma_I-\sigma_{II}}{\Theta+\sigma_I}\left[3\Omega^2-{\cal E}_{I}-Y_T-\frac{1}{3}(\Theta+\sigma_I)(\Theta+\sigma_{II})\right],
\label{une2}
\end{equation}
and
\begin{equation}
{\cal E}_{II}-{\cal E}_{I}=\frac{\sigma_I-\sigma_{II}}{\Theta+\sigma_{II}}\left[3\Omega^2-{\cal E}_{II}-Y_T-\frac{1}{3}(\Theta+\sigma_I)(\Theta+\sigma_{II})\right],
\label{une3}
\end{equation}
where (\ref{theta}) has also been used.

In the above expressions,  ${\cal E}_{I, II}$, are two of the three scalar functions defining the electric part of the Weyl tensor (see \cite{1} for details) (the third scalar function is  denoted ${\cal E}_{KL}$ ).  Also,  $Y_T$, is one  of the structure scalars obtained from the orthogonal splitting of the Riemann tensor which are  defined in eqs.(38-50) in \cite{1}.The others are denoted by  $Y_{I, II, KL}, X_{T, I, II, KL}, Z_{I, II, III, IV}$.

From the above equations and (\ref{sum1}) it follows at once that  as $r\approx 0$
\begin{equation}
{\cal E}_{I}-{\cal E}_{II}=\sum_{n\geq 4}\left[{\cal E}_I^{(n)}(t,\theta)-{\cal E}_{II}^{(n)}(t,\theta) \right]r^{n},
\label{sum5}
\end{equation}
and (\ref{ref1}), unless $\Theta+\sigma_I=\Theta+\sigma_{II}=0$.

However, this last condition cannot be satisfied. Indeed the absence of singularities in $H_2$ at $r=0$ requires from (\ref{ecc7Y}), that $\sigma_{I,II}\approx r$ as $r\approx 0$ which  would produce  $ \Theta\approx 0$, implying in turn  because of (\ref{ecc1Y})
\begin{equation}
Y_T({r=0})=(\mu+3P)_{r=0}=0.
\label{nmnb}
\end{equation}

Now, in order to satisfy (\ref{nmnb}) we have to assume,   the equation of state
\begin{eqnarray}
(\mu+3P)=0,
\label{nm2}
\end{eqnarray}
at least in the neighborhood of  the center.  Excluding this  possibility on physical grounds, we have to assume $\Theta+\sigma_{I,II}\neq0 $.

Next, from the combination of (\ref{dB10Y}),  (\ref{dB12}) and  (\ref{dB13}) we obtain

\begin{equation}
\frac{\tilde G}{\sqrt{B^2r^2+\tilde G^2}}\left[\frac{H_2 \tilde G^\prime}{2 B \sqrt{B^2r^2+\tilde G^2}}+H_1 \left(\frac{\dot B}{B}-\frac{\dot C}{C}\right)\right]=0,
\label{45Y}
\end{equation}
where   $H_1, H_2$ are the two scalar functions defining the magnetic part of the Weyl tensor.

From the above equation it follows that either $\tilde G=0\Rightarrow \Omega=0$, or
\begin{equation}
H_2\Omega+H_1\left[\frac{2\sigma_I+\sigma_{II}}{3}\right]=0,
\label{sum7}
\end{equation}
where (\ref{omegaa})--(\ref{sigmasa}) have been used.

Since $\Omega$ goes to zero at the center, as $r$,  we are left with three possibilities to proceed further, namely: 
\begin{enumerate}
\item The vorticity is assumed to vanish ($\tilde G=0$).
\item $\tilde G\neq0$ and the term within the square bracket in (\ref{sum7}) does not vanish at  the center, meaning that we may write
\begin{equation}
\sigma_I=\sum_{n=0}{\sigma^{(n)}_{I}r^n},\;\; \sigma_{II}=\sum_{n=0}{\sigma^{(n)}_{II}r^n}.
\label{nmn}
\end{equation}

\item $\tilde G\neq0$ and the term within the square bracket in (\ref{sum7}) does vanish at  the center implying that  
\begin{equation}
\sigma_I=\sum_{n\geq1}{\sigma^{(n)}_{I}r^n},\;\;\sigma_{II}=\sum_{n\geq1}{\sigma^{(n)}_{II}r^n}.
\label{nmn1}
\end{equation}
\end{enumerate}

Next, contracting (A.7) in \cite{1} with {\bf K} and {\bf L} we obtain, respectively

\begin{equation}
P^\prime=0\,\,\, \Rightarrow \,\,\, P=P(t,\theta),\label{p1}
\end{equation}

\begin{equation}
\tilde G P_{,t}+P_{,\theta}=0,\label{p2}
\end{equation}

\noindent which, due to the regularity conditions on the axis of symmetry, implies that either $\tilde G=0$ and $P=P(t)$, or $\tilde G\neq0$ and $P=constant$.

\noindent We shall  now analyze the three possible cases mentioned before.
\subsection{$\tilde G=0$}
From the well established link between radiation and vorticity (see \cite{vr} and references therein),  it might be inferred that no gravitational radiation is expected to be emitted in this case.

In what follows we shall provide a formal proof of this result.

As mentioned before, in  this case we have $P=P(t)$; then, from (\ref{Thetaa}), (\ref{primers}) and (\ref{sigmas}) we get
\begin{equation}
\Theta=2\frac{\dot B}{B}+\frac{\dot C}{C},\;\sigma_I=\sigma_{II}=\tilde\sigma=\frac{\dot B}{B}-\frac{\dot C}{C}\label{sigmas'},
\end{equation}
implying:
\begin{equation}
\Theta=2\tilde \sigma+\frac{3\dot C}{C}=\frac{3\dot B}{B}-\tilde \sigma.
\label{nthet}
\end{equation}
Next,  equation  (\ref{ecc1Y}) is the Raychaudhury equation for this case, which reads

\begin{equation}
\dot \Theta +\frac{1}{3}\Theta^2+\frac{2}{3}\tilde \sigma^2+Y_T=0,
\label{nray}
\end{equation}
or, using (\ref{nthet})
\begin{equation}
\dot {\tilde \sigma}-\frac{1}{3}\tilde \sigma ^2+\frac{2}{3}\Theta \tilde \sigma +\frac{3}{2}\frac{\ddot C}{C}+\frac{Y_{T}}{2}=0,
\label{nnray}
\end{equation}
whereas from (\ref{ecc2Y})--(\ref{ecc3Y}), we obtain, respectively
\begin{equation}
\dot {\tilde \sigma}-\frac{1}{3}\tilde \sigma ^2+\frac{2}{3}\Theta \tilde \sigma +Y_I=0,\label{ec2}
\end{equation}

\begin{equation}
Y_{KL}=0 \Rightarrow  X_{KL}=0,\label{ec2bis}
\end{equation}
and 

\begin{equation}
\dot {\tilde \sigma}-\frac{1}{3}\tilde \sigma ^2+\frac{2}{3}\Theta \tilde \sigma +Y_{II}=0.
\label{ec3}
\end{equation}

\noindent Then, from (\ref{nnray})--(\ref{ec3}) we get
\begin{equation}
Y_I=Y_{II}=Y=\frac{3}{2}\frac{\ddot C}{C}+\frac{Y_{T}}{2},
\label{Y}
\end{equation}
and using (42)--(45) and (47)--(50) in \cite{1} it follows that 
\begin{eqnarray}
X_I&=&X_{II}=X, \; \mathcal E_{I}=\mathcal E_{II}=\mathcal E, \; Y=-X=\mathcal E,\nonumber \\ \mathcal E_{KL}&=&X_{KL}=Y_{KL}=0.
\label{XE}
\end{eqnarray}

\noindent Two constraint equations follow from (\ref{ecc5Y}) and (\ref{ecc6Y})
\begin{equation}
2\Theta ^{\prime}-\tilde \sigma ^\prime -3\tilde \sigma \frac{C^\prime}{C}=0,
\label{ec4}
\end{equation}

\begin{equation}
2\Theta _{,\theta}-\tilde \sigma _{,\theta} -3\tilde \sigma \frac{C_{,\theta}}{C}=0,\label{ec5}
\end{equation}
whereas from (\ref{ecc8Y}) and (\ref{ecc7Y}) we get 
\begin{equation}
H_1=-\frac{\tilde \sigma}{2Br}\left (\frac{C_{,\theta}}{C}+\frac{\tilde \sigma_{,\theta}}{\tilde \sigma}\right )=-\frac{(\tilde\sigma C)_{,\theta}}{2BrC},\label{ec6}
\end{equation}

\begin{equation}
H_2=\frac{\tilde \sigma }{2B}\left ( \frac{\tilde \sigma ^\prime}{\tilde \sigma}+\frac{C^\prime}{C}\right)=\frac{(\tilde \sigma C)^\prime}{2BC}.\label{ec7}
\end{equation}

From the two equations above we find that if the magnetic part of the Weyl tensor vanishes, then 

\begin{equation}
\tilde \sigma C=\psi(t),
\label{nu1}
\end{equation}
where $\psi(t)$ is an arbitrary integration function. The above equation implies, because of the regularity condition $C(t,0,\theta)=0$, that $\tilde \sigma=0$.

This last result in turn implies (as mentioned before) that our line element (\ref{1b}) (or (\ref{metric2})) is not the more general one, since in the spherically symmetric limit, it does not contain the Lemaitre-Tolman-Bondi metric. Also, the Szekeres metric  \cite{1S, 2S, 7bis, 8}(its axially symmetric version) cannot be recovered from (\ref{metric2}). In other words, the shear of  the fluid (in our models) is sourced  by the magnetic part of the Weyl tensor. Therefore in the spherically symmetric  limit, we recover the shear--free case.

Another important conclusion emerges from (\ref{ec6}), (\ref{ec7}). Indeed, since $B$ is regular at the origin ($r\approx 0$), whereas $C$ behaves as $C\approx r$, then for  the magnetic Weyl tensor to be regular at the origin, we must demand there $\tilde \sigma \approx r$ (at least). But then, (\ref{ec2}) or (\ref{ec3}) implies that ${\cal E}\approx r$ at the origin.

Also,  from (\ref{ec5}), it follows that  $\Theta_{,\theta} \approx 0$ (in the neighborhood of the  center), and then from (\ref{nthet}) it follows that
\begin{equation}
 \frac{\gamma_{,\theta}}{\gamma}=l(\theta),
\label{nuev2}
\end{equation}
where $l(\theta)$ is an arbitrary function of its argument.

Next, using (\ref{ec4}) and (\ref{ec5}) we may write
\begin{equation}
2(\Theta+\sigma)^{\prime}=\frac{(3\tilde \sigma C)^\prime}{C}
\Rightarrow 2\left(\frac{\dot B}{B}\right)^{\prime}=\frac{(\tilde \sigma C)^\prime}{C},
\label{ntex1}
\end{equation}
\begin{equation}
2(\Theta+\sigma)_{,\theta}=\frac{(3\tilde \sigma C)_{,\theta}}{C}
\Rightarrow 2\left(\frac{\dot B}{B}\right)_{,\theta}=\frac{(\tilde \sigma C)_{,\theta}}{C}.
\label{ntex2}
\end{equation}
Feeding back the two equations above into (\ref{ec6}) and (\ref{ec7}) we obtain
\begin{equation}
BrH_1=-\left(\frac{\dot B}{B}\right)_{,\theta},\;\; BH_2=\left(\frac{\dot B}{B}\right)^\prime.
\label{nu3}
\end{equation}

On the other hand (\ref{dB11}) and (\ref{48Y}), produce, respectively
\begin{equation}
H_{2,\theta}+H_2\left ( \frac{ 2C_{,\theta}}{C}-\frac{B_{,\theta}}{B}\right )=r\left[H^{\prime}_{1}+H_1\left (\frac{2C^\prime}{C}-\frac{(Br)^\prime}{Br}\right )\right ],\label{ecH2}
\end{equation} 
and
\begin{equation}
(H_1BrC^2)^\prime+(H_2BC^2)_{,\theta}=0,
\end{equation}
whereas the combination of  the  equations above with (\ref{ec6})  and (\ref{ec7}) produces
\begin{equation}
H_1C^\prime r+H_2C_{,\theta}=0,
\label{nu4}
\end{equation}
or
\begin{equation}
-\left(\frac{\dot B}{B}\right)_{,\theta} C^\prime+\left(\frac{\dot B}{B}\right)^\prime C_{,\theta}=0.
\label{nu4bis}
\end{equation}
Next, using (\ref{sigmas'}) in (\ref{ntex1}) and (\ref{ntex2}), it follows that 
\begin{equation}
\left(\frac{\dot B}{B}\right)_{,\theta}=\frac{\dot B C_{,\theta}}{BC}-\frac{\dot C_{,\theta}}{C},
\label{nu5}
\end{equation}
\begin{equation}
\left(\frac{\dot B}{B}\right)^\prime=\frac{\dot B C^\prime}{BC}-\frac{\dot C^\prime}{C}.
\label{nu6}
\end{equation}
Now, combining the two equations above with (\ref{nu4bis}) we may write
\begin{equation}
\frac{\dot C_{,\theta}}{C_{,\theta}}=\frac{\dot C^\prime}{C^\prime}\Rightarrow C^\prime=\gamma (r,\theta)C_{,\theta},
\label{nu7}
\end{equation}
implying
\begin{equation}
C=\kappa(t) \tilde C(r,\theta),
\label{nu8}
\end{equation}
and, because of (\ref{nu4})
\begin{equation}
H_2=-r\gamma(r,\theta)H_1.
\label{nu9}
\end{equation}

Then, using (\ref{nu7}) in (\ref{nu4bis}) we get

\begin{equation}
\frac{B^\prime}{B}=\frac{B_{,\theta}}{B}\gamma(r,\theta)+\epsilon(r,\theta),
\label{nu10}
\end{equation}
with
\begin{equation}
\epsilon=\frac{\gamma_{,\theta}r+\gamma \gamma^\prime r^3-1}{r(1+r^2\gamma^2)},
\label{nv1}
\end{equation}

implying
\begin{equation}
B=T(t)\tilde B (r,\theta)\Rightarrow  \frac{\dot B}{B}=\frac{\dot T}{T}=f(t),
\label{nu11}
\end{equation}
where $T(t)$ and $\tilde B (r,\theta)$ are arbitrary functions of their arguments.

Using the above result in (\ref{nu3}), it follows at once that $H_1=H_2=0$, which as mentioned before implies $\tilde \sigma=0$. This last result in turn implies  that the expansion scalar only depends on $t$ and taking into account that $Y_T=4\pi(\mu +P)$, the Raychaudhury equation requires $\mu=\mu(t)$.

Next, (\ref{46Y}) reads in this case
\begin{equation}
\frac{1}{3}\mathcal{E} ^\prime+\frac{C^\prime}{C}\mathcal{E}=0,\label{ec8}
\end{equation}
which due to the fact that $\mathcal{E}\approx 0$ at the origin, implies $\mathcal{E}=0$.

Thus our spacetime is conformally flat, shear--free, and due to the fact that the fluid  is perfect, it is a FRW spacetime, in agreement with the result obtained in \cite{shearfree}.

For the sake of completeness we shall  sketch another proof of the above result in the Appendix B.

Now, from the fact that our system is conformally flat it appears that it does not radiate gravitationally (according to the criterium commented in the Introduction).

Indeed, the super-Poynting vector can be written  (ec.(55) in \cite{1}) as 
\begin{equation}
P_\alpha=P_IK_\alpha+P_{II}L_\alpha,
\label{sp1}
\end{equation}

\noindent where according to  (56) in \cite{1}, and (\ref{XE}), we have  for the two scalars defining the super-Poynting vector
\begin{equation}
P_I=2H_2Y;\;\; P_{II}=-2H_1Y.
\label{sp}
\end{equation}
Thus, the vanishing  of  ${\cal E}$, $H_1$ and $H_2$, as in our case, implies the vanishing of gravitational radiation.
\\
\subsection{$\tilde G\neq0$ and the term within the square bracket in (\ref{sum7}) does not vanish at  the center.}

In this case,  from the regularity of $H_2$ at the center we may write, in the neighborhood of $r\approx 0$,
\begin{equation}
H_2=\sum_{n=0}{H^{(n)}_2r^n},
\label{sum8}
\end{equation}
in which case, (\ref{sum7}) implies 
\begin{equation}
H_1=\sum_{n\geq1}{H^{(n)}_1r^n}.
\label{sum9}
\end{equation}
Replacing the two above expressions in a combination of  (\ref{dB10Y}) and (\ref{dB12}), and using (\ref{sum1}), (\ref{eccB3Y}), (\ref{sum5}), (\ref{ref1}) we find that the lowest order of $r$ in $H_1$ and $H_2$ is raised as
\begin{equation}
H_1=\sum_{n\geq6}{H^{(n)}_1r^n},\;\; H_2=\sum_{n\geq5}{H^{(n)}_2r^n}.
\label{sum8bis}
\end{equation}

Next, from (\ref{46Y})
we may write (close to the center)
\begin{equation}
{\cal E}_I=\sum_{n\geq1}{{\cal E}^{(n)}_Ir^n},\;\; {\cal E}_{II}=\sum_{n\geq1}{{\cal E}^{(n)}_{II}r^n},
\label{sum9bis}
\end{equation}

\noindent for otherwise there would be an inadmissible singularity in the $r$-derivative of the energy density at the origin.

Feeding back the two expressions above  in a combination of (\ref{dB12}) and (\ref{dB13}), it follows from the lowest order in $r$, that  in the neighborhood of the center
\begin{equation}
(\mu+P)\sigma_I\approx O(r),
\label{sum10}
\end{equation}
which implies that, 

\begin{equation}
(\mu+P)\approx O(r),
\label{nm}
\end{equation}
which of course is impossible unless we assume, close to the center,  the equation of state
\begin{eqnarray}
(\mu+P)=0.
\label{nm2}
\end{eqnarray}
 Excluding this   possible situation from physical considerations, we have to require that  
\begin{equation}
\sigma_I=\sum_{n\geq1}{\sigma^{(n)}_I r^n}
\label{nm}
\end{equation}
which in turn implies 
\begin{equation}
\sigma_{II}=\sum_{n\geq1}{\sigma^{(n)}_{II} r^n}
\label{nm}
\end{equation}
because of (\ref{ref1}). 

But of course this contradicts the main assumption of this case about the nonvanishing of the term within the square bracket in (\ref{sum7}), (Eq. (\ref{nmn})).

Thus we have to assume (\ref{nmn1}), implying  that the term within the square bracket in (\ref{sum7}) vanishes at the center as $r$.
\subsection{$\tilde G\neq0$, and the term within the square bracket in (\ref{sum7}) does vanish at  the center.}
Then, it follows at once from (\ref{Thetaa}) and (\ref{sigmaIa}), close to the center, that
\begin{equation}
\frac{\dot B}{B}\approx \frac{\dot C}{C},\;\; \Theta\approx \frac{3\dot C}{C}\approx \frac{3\dot B}{B}.
\label{p31}
\end{equation}

Next, from  the lowest order of  $r$ in (\ref{ecc2Y}) and (\ref{ecc3Y}) it appears  that
\begin{equation}
{\cal E}_1=\sum_{n\geq1}{{\cal E}^{(n)}_Ir^n},\;\;
{\cal E}_{II}=\sum_{n\geq1}{{\cal E}^{(n)}_{II}r^n},
\label{p32}
\end{equation}
and, (\ref{sigmasa}), (\ref{eccB3Y}), (\ref{une2}) and (\ref{une3}), as $r\approx 0$,  produce (\ref{ref1}), (\ref{sum5}) and  
\begin{equation}
{\cal E}_{KL}=\sum_{n\geq5}{{\cal E}_{KL}^{(n)}r^n}.
\label{sum5n}
\end{equation}

Excluding singularities of the scalars $H_1, H_2$, at the origin, we may write:
\begin{equation}
H_1=\sum_{n=0}{H^{(n)}_1r^n}, \;\;H_2=\sum_{n=0}{H^{(n)}_2r^n},
\label{sum8bisb}
\end{equation}

Then, looking for the lowest order of $r$ in (\ref{dB10Y})--(\ref{dB13}) and (\ref{48Y}) we obtain respectively:
\begin{equation}
H^{(0)}_{1,\theta}=-H^{(0)}_{1}\frac{\gamma_{,\theta}}{\gamma},
\label{npul1}
\end{equation}
\begin{equation}
H^{(0)}_{2,\theta}+H^{(0)}_{2}(\frac{2\gamma_{,\theta}}{\gamma}-\frac{B_{,\theta}}{B})-H^{(0)}_{1}=0,
\label{npul2}
\end{equation}
\begin{equation}
H^{(0)}_{2}=-H^{(0)}_{1}(\frac{\gamma_{,\theta}}{\gamma}-\frac{B_{,\theta}}{B}),
\label{npul3}
\end{equation}
\begin{equation}
H^{(0)}_{1,\theta}+H^{(0)}_{1}\frac{B_{,\theta}}{B}-H^{(0)}_{2}=0,
\label{npul4}
\end{equation}

\begin{equation}
3H^{(0)}_{1}+H^{(0)}_{2,\theta}+H^{(0)}_{2}(\frac{2\gamma_{,\theta}}{\gamma}+\frac{B_{,\theta}}{B})=0.
\label{npul5}
\end{equation}

Next, from the lowest order in (\ref{ecc6Y}) it follows that in the neighborhood of the center
\begin{equation}
\Theta_{,\theta} \approx 0 \rightarrow \left(\frac{\dot C}{C}\right)_{,\theta}\approx \left(\frac{\dot B}{B}\right)_{,\theta}\approx 0,
\label{npul6}
\end{equation}
implying right there
\begin{equation}
\gamma=f(t)\alpha(\theta),\;\; B=g(t)\beta(\theta).
\label{npul7}
\end{equation}

Then integrating (\ref{npul1}) we obtain
\begin{equation}
H^{(0)}_{1}=\frac{x(t)}{\alpha(\theta)},
\label{npul8}
\end{equation}
where $x(t)$ is an integration function.

An equation derived from the combination of (\ref{npul2}) and (\ref{npul5}), can also be integrated to produce
\begin{equation}
H^{(0)}_{2}=\frac{y(t)\beta^{1/2}(\theta)}{\alpha^2(\theta)},
\label{npul9}
\end{equation}
and a combination of (\ref{npul2}) and (\ref{npul5}) also produces
\begin{equation}
2H^{(0)}_{1}+H^{(0)}_{2}\frac{B_{,\theta}}{B}=0.
\label{npul9}
\end{equation}
From the equations above it follows that
\begin{equation}
-2\frac{x(t)}{y(t)}=\frac{\beta_{,\theta}}{\alpha \beta^{1/2}}=constant.
\label{npul9}
\end{equation}

However, this last equation  cannot be satisfied. Indeed, because of the reflection symmetry, we have that $\beta(0)=\beta(\pi)$, implying that $\beta_{,\theta}$  must have a change of sign in the interval $[0, \pi]$, whereas $\alpha$ and $\beta$ are positive defined. 

Thus we must put $H^{(0)}_{1}=H^{(0)}_{2}=0$. Accordingly we have:
\begin{equation}
H_2=\sum_{n\geq1}{H^{(n)}_2r^n},
\label{sum8bisbb}
\end{equation}

\begin{equation}
H_1=\sum_{n\geq1}{H^{(n)}_1r^n}.
\label{sumb8b}
\end{equation}
Next, multiplying (\ref{50Y}) by $2$ and subtracting from (\ref{46Y}), we obtain at  the lowest order  of $r$ 
\begin{equation}
{\cal E}_1=\sum_{n\geq2}{{\cal E}^{(n)}_Ir^n},\;\;
{\cal E}_{II}=\sum_{n\geq2}{{\cal E}^{(n)}_{II}r^n}.
\label{p32b}
\end{equation}

Using the expressions above, we are now looking for the lowest order of $r$ in 
(\ref{dB10Y})--(\ref{dB13}) and (\ref{48Y}), we obtain, respectively
\begin{equation}
H^{(1)}_{1,\theta}=-H^{(1)}_{1}\frac{\gamma_{,\theta}}{\gamma},
\label{npul1b}
\end{equation}
\begin{equation}
2H^{(1)}_1-H^{(1)}_{2,\theta}-H^{(1)}_{2}(\frac{2\gamma_{,\theta}}{\gamma}-\frac{B_{,\theta}}{B})=0,
\label{npul2b}
\end{equation}
\begin{equation}
2H^{(1)}_{2}=-H^{(1)}_{1}(\frac{\gamma_{,\theta}}{\gamma}-\frac{B_{,\theta}}{B}),
\label{npul3b}
\end{equation}
\begin{equation}
H^{(1)}_{1,\theta}+H^{(1)}_{1}\frac{B_{,\theta}}{B}-2H^{(1)}_{2}=0,
\label{npul4b}
\end{equation}

\begin{equation}
4H^{(1)}_{1}-H^{(1)}_{2,\theta}-H^{(1)}_{2}(\frac{2\gamma_{,\theta}}{\gamma}+\frac{B_{,\theta}}{B})=0.
\label{npul5b}
\end{equation}

Then, proceeding exactly as we did before,  using (\ref{npul7}), we are lead to 
\begin{equation}
\frac{\beta_{,\theta}\beta^2}{\alpha}=constant,
\label{npul9b}
\end{equation}
which cannot be satisfied  because of the reflection symmetry, as argued before.

Therefore we must put
\begin{equation}
H_2=\sum_{n\geq2}{H^{(n)}_2r^n},
\label{sum8bisbbb}
\end{equation}

\begin{equation}
H_1=\sum_{n\geq2}{H^{(n)}_1r^n}.
\label{sumb8bb}
\end{equation}

At this point we have to stop the procedure followed so far with equations (\ref{dB10Y})--(\ref{dB13}) and (\ref{48Y}), since now the lowest order in $r$ in these equations, contains terms not including $H_1$ and $H_2$. 

Thus let us turn to equations (\ref{ecc5Y}), (\ref{ecc8Y}) (\ref{ecc7Y}).

From the lowest order in  (\ref{ecc8Y})  and  (\ref{ecc7Y}) we find, respectively
\begin{equation}
\sigma^{(1)}_I=\sigma^{(1)}_{II}=\frac{p(t)}{\alpha},
\label{rec1}
\end{equation}
and 
\begin{equation}
\Omega^{(1)}=q(t,\theta)\alpha,
\label{rec2}
\end{equation}
with
\begin{equation}
q_{,\theta}=\frac{2p(t)}{\alpha^2},
\label{rec3}
\end{equation}
where $p(t)$ and $q(t,\theta)$ are arbitrary functions.

Now, by the same arguments based on  the reflection symmetry exposed before, it is easily concluded that $p(t)=0$ implying $\sigma^{(1)}_I=\sigma^{(1)}_{II}=0$. Using this result in the lowest order of $r$ in (\ref{ecc5Y}) we obtain $\Omega^{(1)}=0$. We can now  feed these results back into (\ref{ecc8Y})  and  (\ref{ecc7Y}), and look for the lowest order in $r$. We obtain then that 
\begin{equation}
\sigma_{II}=\sum_{n\geq3}{\sigma^{(n)}_{II}r^n},\;\;\sigma_{I}=\sum_{n\geq3}{\sigma^{(n)}_{I}r^n}.
\label{rec4}
\end{equation}
Using this last result again in (\ref{ecc5Y}), the lowest order in $r$ now implies $\Omega^{(2)}=0$, which in turn, because of (\ref{ecc2Y}) and (\ref{ecc3Y}) implies 
\begin{equation}
{\cal E}_1=\sum_{n\geq3}{{\cal E}^{(n)}_Ir^n},\;\;
{\cal E}_{II}=\sum_{n\geq3}{{\cal E}^{(n)}_{II}r^n}.
\label{p32}
\end{equation}

Using the results above we can now return to (\ref{dB10Y})--(\ref{dB13}) and (\ref{48Y}), since now the lowest  order in $r$, in these equations,  only contains terms with $H_1$ and $H_2$. Doing so we shall raise the lowest order in $r$ of  $H_1$ and $H_2$, until the moment when the lowest order in these equations contains  terms without  $H_1$ and $H_2$. Then we can  go again through the whole cycle above.  Now, it is a simple matter to see that this procedure may be continued as many times as desired, to obtain that 
$H^{(n)}_1=H^{(n)}_2={\cal E}^{(n)}_{I}={\cal E}^{(n)}_{II}=\sigma^{n}_1=\sigma^{n}_{II}=\Omega^{(n)}=0$ for any value of $n\geq 0$, implying in turn that at the center, these quantities as well as their $r$-derivatives of any order vanish. 

Then, assuming that all relevant variables are  of class $C^{\omega}$, i.e. that they  equal their Taylor series expansion around  the center,   we can analytically continue the zero value at the center to the whole configuration and therefore, we obtain a  conformally flat and shear--free spacetime (i.e. F.R.W.).

\section{conclusions}
We have shown that all possible models compatible with the line element (\ref{metric2}) and a perfect fluid, are FRW, and accordingly non--radiating (gravitationally). This clearly indicates that, both, the geodesic and the non--dissipative conditions, are quite restrictive, when looking  for a source  of gravitational waves. 

Having arrived at this point, the relevant question is: does this result make sense from the physical point of view?

To answer to such a question, let us first remember that, already in the seminal Bondi's paper  on gravitational radiation(see section 6 in\cite{7}), it was clearly stated that, not only   in the case of dust, but also in the absence of dissipation in a perfect fluid, the system is not expected  to radiate (gravitationally) due to the reversibility of the equation of state. The rationale supporting this conjecture is very clear: radiation is an irreversible process, this fact emerges at once  if  absorption is taken into account and/or Sommerfeld type conditions, which eliminate inward traveling waves, are imposed. Therefore,  it is obvious that an entropy generator factor should be present in the description of the source.  But  such a factor is absent in a perfect fluid, and more so in a collisionless dust. In other words, the irreversibility of the  process of emission of gravitational waves, must be reflected in the equation of state through an entropy increasing  (dissipative) factor.

In order to delve deeper into this question, let us  invoke here the tight relationship between radiation and vorticity mentioned before (see the beginning of Sec. II A). 

Now,  the equation (\ref{ecc4Y}) in the general (non--geodesic) case (Eq. (B.5) in \cite{1}),reads

\begin{equation}
\Omega _{,\delta}V^\delta +\frac{1}{3}(2\Theta+\sigma _I+\sigma _{II})\Omega +K^{[\mu}L^{\nu]}a_{\mu;\nu}=0,\label{esc51KL}
\end{equation}
which of course reduces to  (\ref{ecc4Y}) if the four acceleration vanishes. 

From (\ref{ecc4Y}) it follows at once that if at any given time,  the vorticity vanishes, then it vanishes at any other time afterwards.  Thus we should not expect gravitational radiation from a physically meaningful system, radiating for a finite period of time  (in a given time interval), for otherwise such a radiation will not be accompanied by the presence of vorticity. 

But, what happens  for the perfect (non--dissipative, non--geodesic) fluid?

In this latter case, the condition of thermal  equilibrium (absence of dissipative flux) reads (see eq. (57) in \cite{1})
\begin{equation}
a_\mu=-h^\beta_\mu \Gamma_{,\beta},
\label{t1}
\end{equation}
where $\Gamma=\ln T$, and  $T$ denotes the  temperature.

Feeding back  (\ref{t1}) into (\ref{esc51KL}) we get
\begin{equation}
\Omega _{,\delta}V^\delta +\frac{1}{3}(2\Theta+\sigma _I+\sigma _{II}+V^\mu \Gamma_{,\mu})\Omega =0.\label{esc51KLb}
\end{equation}
Thus, even if the fluid is not geodesic, but is non--dissipative, the situation is the same as in the geodesic case, i.e. the vanishing of vorticity at any given time implies its vanishing for any time in the future.

This result is in full agreement with earlier works indicating that vorticity generation is sourced by entropy gradients 
\cite{Croco}--\cite{75}. At the same time we confirm, by invoking the radiation--vorticity link,  the Bondi's conjecture about the absence of radiation for non--dissipative systems.

Finally, two comments are in order before concluding:
\begin{itemize} 
\item In a recent work  \cite{75b}, the role played by magnetic fields in the generation and survival of vorticity, has been brought out. This strongly suggest that the inclusion of magnetic fields in the discussion of gravitationally radiating sources, deserves further attention. 
\item Geodesic fluids not belonging to the class considered here (Szekeres) have also been shown not to produce gravitational radiation \cite{bill}. This strengthens further the case of the non--radiative character of pure dust distributions.

\end{itemize}
\begin{acknowledgments}
L.H. thanks  Departament de F\'isica at the  Universitat de les  Illes Balears, for financial support and hospitality. ADP  acknowledges hospitality of the
 Departament de F\'isica at the  Universitat de les  Illes Balears. J.O. acknowledges financial support from the Spanish
Ministry of Science and Innovation (grant FIS2009-07238).
\end{acknowledgments}

\appendix
\section{Summary of equations for the geodesic case}
Below we shall write the equations of the framework developed in \cite{1}, for the geodesic and perfect fluid case. Then, equations  {\bf B.1}--  {\bf B.18} in \cite{1}, read, respectively

\noindent From {\bf B.1}
\begin{equation}
\dot \Theta+\frac{1}{3}\Theta^2+\frac{2}{9}\left(\sigma_I^2 +\sigma_{I}\sigma_{II}+\sigma_{II}^2\right)+Y_T=2\Omega ^2.
\label{ecc1Y}
\end{equation}

\noindent From {\bf B.2}

\begin{equation}
\dot \sigma_{I}+\frac{1}{9}\sigma_{I}^2+\frac{2}{3}\Theta \sigma_{I} -\frac{2}{9}\sigma_{II}\left(\sigma_I +\sigma_{II}\right)+Y_I=\Omega ^2.
\label{ecc2Y}
\end{equation}
\\
\noindent From {\bf B.3}

\begin{equation}
\frac{1}{3}\left(\sigma_{I}-\sigma_{II}\right)\Omega + Y_{KL}=0.
\label{eccB3Y}
\end{equation}
\noindent From {\bf B.4}

\begin{equation}
\dot \sigma_{II}+\frac{1}{9}\sigma_{II}^2+\frac{2}{3}\Theta \sigma_{II} -\frac{2}{9}\sigma_{I}\left(\sigma_I +\sigma_{II}\right)+Y_{II}=\Omega ^2.
\label{ecc3Y}
\end{equation}
\\
\noindent From {\bf B.5}

\begin{equation}
\dot \Omega +\frac{1}{3}\left(2\Theta+\sigma_{I}+\sigma_{II}\right)\Omega=0.
\label{ecc4Y}
\end{equation}

\begin{widetext}

\noindent From {\bf B.6}
\begin{eqnarray}
-\frac{1}{\sqrt{B^2r^2+\tilde G^2}}\left [ \Omega _{,\theta}+\tilde G \dot\Omega+ \Omega\left(\frac{\tilde G\dot C}{C}+\frac{C_{,\theta}}{C}\right )\right]
+\frac{1}{3B}\left\{2 \Theta^\prime-\sigma_{I}^\prime
-\sigma_{I}\left[\frac{2C^{\prime}}{C}+\frac{\left(B^2r^2+\tilde G^2\right)^{\prime}}{2\left(B^2r^2+\tilde G^2\right)}\right]\right.\nonumber\\
\left.-\sigma_{II}\left[\frac{C^{\prime}}{C}-\frac{\left(B^2r^2+\tilde G^2\right)^{\prime}}{2\left(B^2r^2+\tilde G^2\right)}\right]\right\}=0. \nonumber\\
\label{ecc5Y}
\end{eqnarray}

\noindent From {\bf B.7}
\begin{eqnarray}
\frac{1}{B}\left(\Omega^{\prime}+\Omega \frac{C^{\prime}}{C}\right)
+\frac{1}{3\sqrt{B^2r^2+\tilde G^2}}\left\{\left(2\Theta-\sigma_{II}\right)_{,\theta}+\tilde G \left(2\Theta-\sigma_{II}\right)^{.}\right.\nonumber\\
\left.+\sigma_{I}\left[\frac{B_{,\theta}}{B}-\frac{C_{,\theta}}{C}+\tilde G\left(\frac{\dot B}{B}-\frac{\dot C}{C}\right)\right]-\sigma_{II}\left[\frac{B_{,\theta}}{B}+\frac{2C_{,\theta}}{C}+\tilde G\left(\frac{\dot B}{B}+\frac{2\dot C}{C}\right)\right]\right\}=0.\nonumber\\
\label{ecc6Y}
\end{eqnarray}

\noindent From {\bf B.8}
\begin{eqnarray}
H_1=-\frac{1}{2B}\left\{\Omega^{\prime}-\Omega\left[\frac{C^{\prime}}{C}-\frac{\tilde G \tilde G^{\prime}}{2\left(B^2r^2+\tilde G^2\right)}\right]\right\}
-\frac{1}{6\sqrt{B^2r^2+\tilde G^2}}\left\{\left(2\sigma_{I}+\sigma_{II}\right)_{,\theta}\right.\nonumber\\
\left.+\sigma_{I}\left[\frac{B_{,\theta}}{B}+\frac{C_{,\theta}}{C}-\tilde G\left(\frac{\dot B}{B}-\frac{\dot C}{C}\right)\right]-\sigma_{II}\left[\frac{B_{,\theta}}{B}-\frac{2C_{,\theta}}{C}+\tilde G\left(\frac{2\dot B}{B}-\frac{2\dot C}{C}\right)\right]\right\}.
\nonumber\\
\label{ecc8Y}
\end{eqnarray}
\noindent From {\bf B.9}
\begin{eqnarray}
H_2=+\frac{1}{6B}\left\{\left(\sigma_{I}+2\sigma_{II}\right)^{\prime}+\sigma_{I}\left[\frac{2C^\prime}{C}-\frac{(Br)(Br)^{\prime}}{B^2r^2+\tilde G^2}\right]
+\sigma_{II}\left[\frac{C^\prime}{C}+\frac{2(Br)(Br)^{\prime}+3\tilde G \tilde G^\prime}{2\left(B^2r^2+\tilde G^2\right)}\right]\right\}\nonumber\\
-\frac{1}{2\sqrt{B^2r^2+\tilde G^2}}\left\{ \Omega_{,\theta}-\Omega \left[\frac{C_{,\theta}}{C}+\tilde G\left(\frac{\dot B}{B}+\frac{\dot C}{C}\right)\right]\right\}.
\label{ecc7Y}
\end{eqnarray}
\noindent From {\bf B.10} 
\begin{eqnarray}
\frac{1}{3}\dot{\mathcal E}_{I}+\frac{4\pi}{3}(\mu+P)\sigma_{I}-\Omega \mathcal E_{KL}
+\frac{\mathcal E_{I}}{9}\left(3\Theta+\sigma_{II}-\sigma_{I}\right)+\frac{\mathcal E_{II}}{9}\left(2\sigma_{II}+\sigma_{I}\right)\nonumber\\
-\frac{1}{\sqrt{B^2r^2+\tilde G^2}}\left(H_{1,\theta}+H_1\frac{C_{,\theta}}{C}\right)
-\frac{H_2}{B}\left[
\frac{C^{\prime}}{C}-\frac{2(Br)(Br)^{\prime}+\tilde G \tilde G^\prime}{2\left(B^2r^2+\tilde G^2\right)}\right]
=0.
\label{dB10Y}
\end{eqnarray}

\noindent From {\bf B.11} 
\begin{eqnarray}
2\dot{\mathcal E}_{KL}+\mathcal E_{KL}\left(2\Theta-\sigma_{I}-\sigma_{II}\right)+\frac{\Omega}{3}\left(\mathcal E_{I}-\mathcal E_{II}\right)
+\frac{1}{B}\left\{H_{1}^{\prime}+H_{1}\left[\frac{2C^{\prime}}{C}-\frac{2(Br)(Br)^{\prime}+\tilde G \tilde G^{\prime}}{2\left(B^2r^2+\tilde G^2\right)}\right]\right\}\nonumber\\
-\frac{1}{\sqrt{B^2r^2+\tilde G^2}}\left\{H_{2,\theta}+H_{2}\left[\frac{2C_{,\theta}}{C}-\frac{B_{,\theta}}{B}
-\tilde G\left(\frac{\dot B}{B}-\frac{\dot C}{C}\right)\right]\right\}=0.
\label{dB11}
\end{eqnarray}

\noindent From {\bf B.12} 
\begin{eqnarray}
\frac{1}{3}\dot{\mathcal E}_{II}+\frac{4\pi}{3}(\mu+P)\sigma_{II}+\Omega \mathcal E_{KL}
+\frac{\mathcal E_{II}}{9}\left(3\Theta+\sigma_{I}-\sigma_{II}\right)+\frac{\mathcal E_{I}}{9}\left(2\sigma_{I}+\sigma_{II}\right)\nonumber\\
+\frac{1}{B}\left(H_{2}^{\prime}+H_{2}\frac{C^{\prime}}{C}\right)
+\frac{H_{1}}{\sqrt{B^2r^2+\tilde G^2}}\left[\frac{C_{,\theta}}{C}-\frac{B_{,\theta}}{B}-\tilde G\left(\frac{\dot B}{B}-\frac{\dot C}{C}\right)\right]=0.\nonumber\\
\label{dB12}
\end{eqnarray}

\noindent From {\bf B.13} 
\begin{eqnarray}
-\frac{1}{3}\left(\mathcal E_{I}+\mathcal E_{II}\right)^{.}-\frac{1}{3}(\mathcal E_{I}+\mathcal E_{II})\Theta-\frac{4\pi}{3}(\mu+P)(\sigma_{I}+\sigma_{II})
-\frac{\mathcal E_{I}}{9}\left(2\sigma_{II}+\sigma_{I}\right)-\frac{\mathcal E_{II}}{9}\left(2\sigma_{I}+\sigma_{II}\right)\nonumber\\
+\frac{1}{\sqrt{B^2r^2+\tilde G^2}}\left(H_{1,\theta}+H_{1}\frac{B_{,\theta}}{B}\right)
-\frac{1}{B}\left\{H_{2}^{\prime}+H_{2}\left[\frac{(Br)(Br)^{\prime}+\tilde G \tilde G^{\prime}}{B^2r^2+\tilde G^2}\right]\right\}=0.\nonumber\\
\label{dB13}
\end{eqnarray}
\noindent From {\bf B.14}
\begin{eqnarray}
\frac{1}{3B}\left\{\mathcal{E}_I ^\prime+\mathcal{E}_I \left[\frac{2C^\prime}{C}+\frac{(B^2r^2+\tilde G^2)^\prime}{2(B^2r^2+\tilde G^2)}\right]
+\mathcal{E}_{II} \left[\frac{C^\prime}{C}-\frac{(B^2r^2+\tilde G^2)^\prime}{2(B^2r^2+\tilde G^2)}\right]\right\}\nonumber\\
+\frac{1}{\sqrt{B^2r^2+\tilde G^2}}\left\{\mathcal E_{KL,\theta}+\tilde G\dot {\mathcal E}_{KL}
+\mathcal E_{KL}\left[\frac{2 B_{,\theta}}{B}+\frac{C_{,\theta}}{C}+\tilde G\left(\frac{2\dot B}{B}+\frac{\dot C}{C}\right)\right]\right\}\nonumber\\
-\frac{1}{3}H_2(\sigma_{I}+2\sigma_{II}) -3\Omega H_1=\frac{8\pi}{3B}\mu ^\prime.\nonumber\\
\label{46Y}
\end{eqnarray}
\noindent From {\bf B.15}
\begin{eqnarray}
\frac{1}{B}\left\{\mathcal{E}_{KL} ^\prime+\mathcal{E}_{KL} \left[\frac{C^\prime}{C}+\frac{(B^2r^2+\tilde G^2)^\prime}{B^2r^2+\tilde G^2}\right]\right\}
+\frac{1}{3\sqrt{B^2r^2+\tilde G^2}}\left\{\mathcal E_{II,\theta}+\tilde G\dot {\mathcal E_{II}}\right.\nonumber\\
\left.+\mathcal{E}_{I} \left[\frac{C_{,\theta}}{C}-\frac{B_{,\theta}}{B}+\tilde G\left(\frac{\dot C}{C}-\frac{\dot B}{B}\right)\right]
+\mathcal{E}_{II} \left[\frac{2C_{,\theta}}{C}+\frac{B_{,\theta}}{B}+\tilde G\left(\frac{2\dot C}{C}+\frac{\dot B}{B}\right)\right]\right\}
\nonumber\\
+\frac{1}{3}H_1(2\sigma_{I}+\sigma_{II}) -3\Omega H_2=\frac{8\pi}{3\sqrt{B^2r^2+\tilde G^2}}(\tilde G\dot \mu+\mu_{,\theta}).\nonumber\\
\label{47Y}
\end{eqnarray}
\noindent From {\bf B.16}
\begin{eqnarray}
 \frac{1}{3}\mathcal E_{KL}\left(\sigma_{II}-\sigma_{I}\right)-\frac{1}{B}\left\{H_1^\prime + H_1\left[\frac{2C^\prime}{C}
+\frac{(B^2r^2+\tilde G^2)^\prime}{2(B^2r^2+\tilde G^2)}\right]\right\}\nonumber\\
-\frac{1}{\sqrt{B^2r^2+\tilde G^2}}\left\{H_{2,\theta}+\tilde G \dot H_{2}+H_2 \left[\frac{B_{,\theta}}{B}+\frac{2C_{,\theta}}{C}+\tilde G\left(\frac{\dot B}{B}+\frac{2 \dot C}{C}\right)\right]\right\}
=\left[8\pi(\mu+P)-(\mathcal{E}_I+\mathcal E_{II})\right]\Omega.\nonumber\\
\label{48Y}
\end{eqnarray}
\noindent From {\bf B.17}
\begin{eqnarray}
 \frac{1}{3 \sqrt{B^2r^2+\tilde G^2}}\left\{\left(\mathcal E_{I}+\mathcal E_{II}\right)_{,\theta}+\tilde G \left(2\mathcal E_{II}+\mathcal E_{I}\right)^{.}
+\mathcal E_{I}\left[\frac{C_{,\theta}}{C}+\tilde G \left(\frac{\dot C}{C}+ \frac{B^2r^2}{B^2r^2 + \tilde G^2}\frac{\dot B}{B}\right)\right]\right.\nonumber\\
\left.+\mathcal E_{II}\left[\frac{2C_{,\theta}}{C}+2\tilde G \left(\frac{\dot C}{C}+ \frac{B^2r^2}{B^2r^2 + \tilde G^2}\frac{\dot B}{B}\right)\right]\right\}
+\frac{\mathcal E_{KL}}{B}\left[\frac{C^\prime}{C}+\frac{\tilde G \tilde G^\prime}{2(B^2r^2+\tilde G^2)}\right]-\dot H_1 
\nonumber\\
-\frac{1}{3}H_1 (3\Theta +\sigma_{II}-\sigma_{I})-H_2\Omega = \frac{4\pi}{3  \sqrt{B^2r^2+\tilde G^2}}\mu_{,\theta}.
\label{49Y}
\end{eqnarray}

\noindent From {\bf B.18}
\begin{eqnarray}
\frac{1}{3B}\left\{\left(\mathcal E_{I}+\mathcal E_{II}\right)^\prime+\mathcal E_{I}\left(\frac{2C^\prime}{C}+\frac{\tilde G \tilde G^\prime}{B^2r^2+\tilde G^2}\right)+\mathcal E_{II}\left[\frac{C^\prime}{C}+\frac{\tilde G \tilde G^\prime}{2(B^2r^2+\tilde G^2)}\right]\right\}\nonumber\\
+ \frac{1}{\sqrt{B^2r^2+\tilde G^2}}\left\{\tilde G \dot{\mathcal E}_{KL}+\mathcal E_{KL}\left[\frac{C_{,\theta}}{C}+\tilde G\left(\frac{\dot B}{B}+\frac{\dot C}{C}\right)\right]\right\}
+\dot H_2 +\frac{1}{3}H_2(3\Theta +\sigma_{I}-\sigma_{II})-H_1\Omega  =\frac{4\pi}{3B}\mu ^\prime.\nonumber\\ \nonumber\\
\label{50Y}
\end{eqnarray}
\end{widetext}

\section{Conformal flatness implied by the vanishing of vorticity}
We shall here provide an alternative proof of the result exhibited in Sec.II about the consequence of assuming vanishing vorticity.

First of all we observe that due to  (\ref{nuev3}), (\ref{nuev2}) and (\ref{nu7}) we may write  at $r\approx 0$
\begin{equation}
\gamma=\frac{L(\theta}{r}, \;\;\; L(\theta)\equiv \frac{1}{l(\theta)},
\label{nuev4}
\end{equation}
where $L(\theta)$ must be a regular function of $\theta$.

Then, using (\ref{nuev4}) in (\ref{nu9}) (always at $r\approx 0$) we obtain
\begin{equation}
H_2=-L(\theta)H_1.
\label{nuev5}
\end{equation}

\noindent Next,   from (\ref{dB10Y}) and (\ref{dB12}) we obtain
\begin{equation}
-H_{1,\theta}-H_1\left ( \frac{2 C_{,\theta}}{C}-\frac{B_{,\theta}}{B}\right )=r\left[H^\prime_{2}+H_2\left (\frac{2C^\prime}{C}-\frac{(Br)^\prime}{Br}\right )\right ].\label{ecH1}
\end{equation}

\noindent Then, introducing the auxiliary function $b(t,r,\theta)$ defined by
\begin{equation}
Br=bC^2,\label{ecb}
\end{equation}
\noindent the equations (\ref{ecH1}) and (\ref{ecH2}) can be rewritten as:
\begin{eqnarray}
-\left(\frac{H_1}{b}\right)_{,\theta}=r\left(\frac{H_2}{b}\right)^\prime,\label{ec11'}
\\
\left(\frac{H_2}{b}\right)_{,\theta}=r\left(\frac{H_1}{b}\right)^\prime,\label{ec12'}
\end{eqnarray}

or
\begin{eqnarray}
\left(\frac{H_1}{b}\right)^{\prime\prime}+\frac{1}{r}\left(\frac{H_1}{b}\right)^\prime+\frac{1}{r^2}\left(\frac{H_1}{b}\right)_{,\theta \theta}
=0,\label{ec11''}
\\
\left(\frac{H_2}{b}\right)^{\prime\prime}+\frac{1}{r}\left(\frac{H_2}{b}\right)^\prime+\frac{1}{r^2}\left(\frac{H_2}{b}\right)_{,\theta\theta}=0.\label{ec12''}
\end{eqnarray}

\noindent The equations (\ref{ec11''}) and (\ref{ec12''}) can be integrated to obtain

\begin{eqnarray}
H_1=b \sum _{n=1} ^\infty r^n \left[\alpha_n (t)\cos(n\theta)+\beta_n(t) \sin(n\theta)\right],
\label{H1}
\\
H_2=b \sum _{n=1} ^\infty r^n\left[- \beta_n (t)\cos(n\theta)+ \alpha_n(t) \sin(n\theta)\right],
\label{H2}
\end{eqnarray}
where coefficients $\alpha_n (t), \beta_n(t)$ are arbitrary functions of $t$.

Observe  that for $n=0$ we have 

\begin{equation}
H_1=\frac{Br}{C^2}\alpha_0(t), \; H_2=-\frac{Br}{C^2}\beta_0(t).
\label{H1n0BC}
\end{equation}
However, in the neighborhood of $r\approx 0$, we have $C\approx r$; implying that $H_1$ and $H_2$ are singular at the origin, and therefore we must put $\alpha_0(t), \beta_0(t)=0$.

Next, feeding back (\ref{H1}) and (\ref{H2}) into (\ref{ecH1}), we have a system of equations for each order of $r^n$. Using  (\ref{nuev5}) we obtain
\begin{equation}
\alpha_1=\beta_1=0,
\label{nuev7}
\end{equation}
where we have explicitly used the fact that $L(\theta)$ is a regular function of $\theta$ (at least on the symmetry axis).

Repeating  the calculations for the consecutive orders, it is a simple matter to check that 
\begin{equation}
\alpha_n=\beta_n=0,\;\; {\rm for \; any \; n},
\label{nuev7b}
\end{equation}
thereby proving that $H_1=H_2=0$.  The remaining of the proof follows as in Sec. III.

\end{document}